\documentclass[prd, preprint,nofootinbib, superscriptaddress]{revtex4-1}    % (no)showpacs:= (no) display PACS number (default = no)
\usepackage{amssymb,bbold}
\usepackage{amsmath}
\usepackage[usenames,dvipsnames]{color}
\usepackage{tensor}
\usepackage{bbm}
\usepackage{graphicx}
\usepackage{cancel}
\usepackage{hyperref}

\hypersetup{
    colorlinks=true,         % false: boxed links; true: colored links, false is default
    linkcolor=blue,          % color of internal links, red is default
    citecolor=red,        % color of links to bibliography, 'green' is default
    urlcolor=Violet             % color of external links, cyan is default
}
\let\oldmarginpar\marginpar
\renewcommand\marginpar[1]{\oldmarginpar{\color{red}\raggedright\scriptsize #1}}
\newcommand{\eq}[1]{(\ref{eq:#1})}
\newcommand{\pb}[2]{\ensuremath{\lf\{#1,#2 \rt\}}}
\newcommand{\mean}[1]{\ensuremath{\lf\langle #1 \rt\rangle }}

\newcommand{\ddiby}[2]{\ensuremath{\frac{\delta #1}{\delta #2}}}

\newcommand{\ind}[2]{\ensuremath{\indices{^{#1}_{#2}}}}
\newcommand{\eps}[2]{\ensuremath{\epsilon\indices{^{#1}_{#2}}}}

\def\lf {\ensuremath{\left}}
\def\rt {\ensuremath{\right}}

\begin{document}

%opening
\title{$2+1$ gravity on the conformal sphere}
\author{Sean Gryb}%\email{sean.gryb@gmail.com}
\affiliation{Institute for Theoretical Physics, Utrecht University,\\Leuvenlaan 4, 3584 CE Utrecht, The Netherlands}
\author{Flavio Mercati}
\affiliation{School of Mathematical Sciences, University of Nottingham,\\ Nottingham NG7 2RD, United Kingdom}
\affiliation{Perimeter Institute for Theoretical Physics,\\ Waterloo, Ontario N2L 2Y5, Canada\\\vspace{1cm}}

\begin{abstract}
    We show that there are 2 equivalent first order descriptions of $2+1$ gravity with non--zero cosmological constant. One is the well--known spacetime description and the other is in terms of evolving conformal geometry. The key tool that links these pictures is Cartan geometry, a generalization of Riemannian geometry that allows for geometries locally modeled off arbitrary homogeneous spaces. The two different interpretations suggest two distinct phase space reductions. The spacetime picture leads to the $2+1$ formulation of General Relativity due to Arnowitt, Deser, and Misner while the conformal picture leads to Shape Dynamics. Cartan geometry thus provides an alternative to symmetry trading for explaining the equivalence of General Relativity and Shape Dynamics.    
    %We show that gravity, with positive cosmological constant in $2+1$ dimensions, can be formulated as a theory of dynamic conformal spatial geometry. Exploiting the isomorphism between the isometry group of de~Sitter space in $D+1$ dimensions and the conformal group in $D$ dimensions, we reinterpret the Chern--Simons formulation of $2+1$ gravity as a gauge theory of a conformal connection. In Cartan's generalization of geometry, this connection represents an evolving spatial geometry locally modeled off the conformal sphere (instead of the flat plane of Riemannian geometry). A suitable phase space reduction gives shape dynamics, a reformulation of general relativity where spacetime foliation invariance is traded for local spatial conformal invariance. This remodeling explains, in $2+1$ dimensions, the remarkable success of the York procedure for solving the initial value problem of general relativity and the uniqueness of the shape dynamics Hamiltonian.
\end{abstract}

\maketitle

\tableofcontents

\section{Introduction}

In the influential essay \emph{La mesure du temps} (\emph{The measure of time}, \cite{Poincare:mesure_du_temps}) from 1889, Poincar\'e raises the question of whether simultaneity is absolute.\footnote{He also makes the less recognized observation that \emph{duration} is relative.} A year later \cite{Poincare:local_time}, he introduces the concept of \emph{local time} to explain the constancy of the speed of light for all observers. By 1905, Poincar\'e and, independently, Einstein had worked out the necessary steps leading to relativity of simultaneity and the birth of special relativity. This work culminated in 1908 with Minkowski's inception of spacetime where ``space by itself, and time by itself, are doomed to fade away into mere shadows, and only a kind of union of the two will preserve an independent reality''\cite{Minkowski:seminal_address}. Strengthened by the development, and phenomenal success, of general relativity (GR), Minkowski's spacetime has been the dominant (if not the only) framework for understanding relativistic phenomena.

But what if history had been different? What if Poincar\'e and Einstein had realized that relativity of spatial \emph{size} -- instead of \emph{simultaneity} -- could be used to explain relativistic phenomena? Quite possibly, they would have discovered some form of \emph{shape dynamics} (SD), a theory that is equivalent to GR in that it reproduces its physical predictions.\footnote{Equivalence is restricted to CMC solutions but is consistent with current observations.} This new theory was developed in \cite{barbour_el_al:physical_dof,gryb:shape_dyn,Gomes:linking_paper} and implements Barbour's idea of relativity of spatial size and Mach's principles \cite{barbour:bm_review,JuliansReview}. In SD, foliation invariance is traded for the local scale invariance. Specifically, this is an invariance of the spatial metric, $g_{ab}$, of the form $g_{ab}(x) \to e^\phi(x) g_{ab}(x)$ called \emph{conformal} (or sometimes \emph{Weyl}) invariance. SD is still a theory of dynamic spatial geometry, in that it is invariant under spatial diffeomorphisms, but it is not a theory of spacetime because it is not invariant under the full set of spacetime diffeomorphisms. Instead, it is a theory of evolving conformal geometry. Thus, the starting ontology and the resulting phase space are the same as that of Arnowitt, Deser, and Misner (ADM) (i.e., spatial metrics and their conjugate momentum densities) and the physical degrees of freedom are also the same. What is different are the symmetries.

An alternative history of this kind would have no doubt unfolded very differently from the one we currently know. Indeed, it is likely that Riemannian geometry itself would not have played a central role since the concept of spacetime would not have been primary. Instead, it is more likely that conformal approaches to geometry, such as \emph{tractor calculus} \cite{Sharpe:Cartan_geometry}, would have been central to the formalism. But is there a simple and elegant description of SD in terms of tractor calculus that could rival the remarkably successful description of GR in terms of Riemannian geometry? This paper is intended as a small first step towards developing such a description.

We will show that it is possible to rewrite the spacetime framework of $2+1$ gravity in terms of dynamic conformal geometry. When interpreted in this way, the Einstein equations can immediately be written as the fundamental equations of SD. Two things make this equivalence possible: 1) the isomorphism between the isometry groups of the homogeneous solutions of the Einstein equations (with non--zero cosmological constant) and the conformal group in one lower dimension, and 2) the time component of a gauge connection is always non--dynamical in any gauge theory. We make use of Cartan's generalization of geometry, which contains both Riemannian geometry -- and, thus, the spacetime picture -- and tractor calculus (or \emph{conformal} Cartan geometry) -- and, thus, the conformal picture. The end result is a precise duality between these two pictures. The Chern--Simons (CS) formulation of $2+1$ gravity is seen to be a gauge theory of a dynamic conformal connection: the conformal Cartan connection. This suggests that, had tractor calculus received as much attention by physicists as Riemannian geometry, our intuition about gravity may have been rooted as much in the conformal approach as in the spacetime approach.\footnote{We have no evidence yet that one approach may be more fundamental than to the other. Clearly, both pictures are valuable and may be better suited for solving different problems.}

This may explain the striking utility of conformal methods in gravity. As early as Dirac \cite{Dirac:CMC_fixing}, conformal transformations were used to fix a foliation in GR. York \cite{York:cotton_tensor,York:york_method_prl} elaborated on these techniques and developed a general method for solving the initial value problem. This technique is still the only known method for solving the initial value problem in full generality for an extensive set of spatial topologies and boundary conditions. In $2+1$ gravity, Moncrief showed \cite{Moncrief:2_plus_1_shape_space} (see also \cite{Carlip:book}), using conformal techniques closely related to York's procedure, that the reduced phase space is a conformally invariant \emph{shape space} (or, more precisely, \emph{Teichm\"uller space}). Furthermore, the conformal decomposition, introduced in \cite{York:york_method_long}, can be used in perturbation theory to show that the physical degrees of freedom of the first order fluctuations of free gravity are conformally invariant.\footnote{More precisely, the metric fluctuations are transverse--traceless with respect to the background metric.} More recently, the uniqueness theorems involved in the York procedure \cite{Niall_73} have been exploited for constructing SD \cite{gryb:shape_dyn,Gomes:linking_paper}.

There are several technical advantages offered by the SD approach. For instance, the constraint algebra of the theory closes under structure constants rather than the structure functions found in GR. Also, the local constraints are linear in the momenta. The implications of these simplifications for the quantum theory are currently being explored and some preliminary results have been obtained. In \cite{Bodendorfer:lqg_without_Ham,Bodendorfer:conf_coupled_sd} for example, Loop Quantum Gravity and SD methods were extended to quantize static spacetimes conformally coupled to a scalar field.\\
On top of these concrete results, there are many promising speculative ideas that make use of conformal invariance in one dimension less than that of the spacetime. Ho\v rava, for instance, has proposed a UV completion of GR where spatial conformal invariance is manifest at high energy \cite{Horava:lif_point}. In string theory, the isomorphism between the isometry group of anti--de~Sitter (AdS) space in $D+1$ dimensions and the conformal group in $(D-1,1)$ dimensions was a key motivation behind Maldacena's AdS/CFT conjecture \cite{Maldacena:ads_cft} and, in particular, the construction of the dictionary \cite{Witten} relating bulk and boundary degrees of freedom. Since the original conjecture, many rigorous results have been obtained and the correspondence has been extended to more general gauge/gravity dualities but a complete understanding of the mechanism behind the correspondence has not been established. Needless to say, conformal methods have proved extremely useful both for producing concrete results in classical general relativity and for providing exciting conjectures about the quantum theory. This is in spite of the fact that the mechanism at work behind these methods has remained a mystery. We propose that this could be deeply connected to the existence of the SD description of gravity and the corresponding relation between evolving conformal geometry and spacetime geometry.
The transition to SD outlined in this paper can easily be extended to the case of negative cosmological constant where the isomorphism between the conformal group and the isometry group of AdS can be exploited, although there are some subtleties discussed in section~\ref{sec:ads cft}. In this context, the reinterpretation of the CS connection as a dynamic conformal geometry suggests a compelling possibility: that the Cartan geometric picture linking spacetime geometry and evolving conformal geometry could lead to a deeper understanding of the mechanism behind the AdS/CFT correspondence.

\subsection{Outline}

The outline of the paper is as follows. We begin with a short description of Cartan geometry in section~\ref{sec:Cartan geometry}, highlighting only the features that we will need to motivate the interpretation of our new variables in terms of conformal geometry. We will then review the CS formulation of $2+1$ gravity in section~\ref{sec:CS} and, in the following two subsections, give a gauge fixing and subsequent phase space reduction leading to the $2+1$ ADM constraints in frame field form. In section~\ref{sec:conf variables}, we rewrite the $2+1$ CS action in terms of evolving conformal geometry and show, in section~\ref{sec:SD phase space}, how this theory suggests a natural gauge fixing and phase space reduction leading to the SD constraints, in frame field form. We end with some comments on how to define the SD Hamiltonian in terms of these new variables and how to extend this work to higher dimensions.

\begin{figure}[b]
\begin{center}
\includegraphics[width=0.45\textwidth]{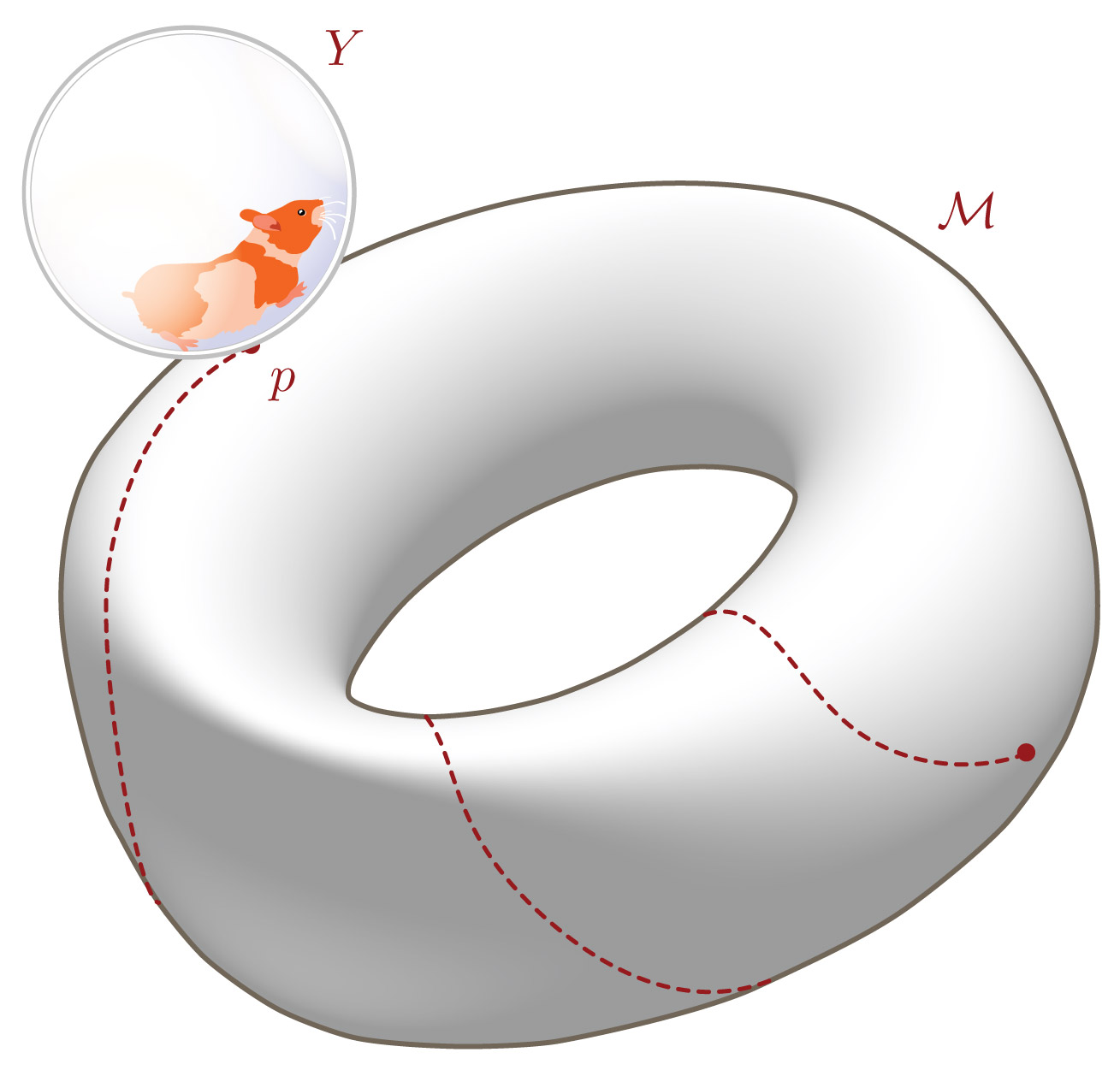}
\caption{Hamster ball rolling over a manifold (thanks to D. Wise for the picture concept).}\label{Hamsterball}
\end{center}
\end{figure}

\section{Cartan geometry}\label{sec:Cartan geometry}

Before presenting our main argument, we will review some elements of Cartan geometry that will be useful for motivating our manipulations. The literature on Cartan geometry and its conformal aspects, described by tractor calculus, is vast. A complete review of this is beyond the scope of this paper. The interested reader is referred to Sharpe's book \cite{Sharpe:Cartan_geometry} for details. For a short pedagogical introduction with applications to physics, see \cite{Wise:MM_paper}.\footnote{For an alternate perspective see \cite{Westman:Cartan_waywisers}.}

Cartan geometry is a generalization of Riemannian geometry where the tangent space of a manifold $\mathcal M$ is locally identified with the tangent space of an arbitrary model homogeneous space so that the local metric structure of the manifold is inherited from the natural metric on the homogeneous space (if such a metric exists). The model space, $Y=G/H$, is given by quotienting a group $G$ by one of its closed subgroups $H$. The structure of the geometry is determined by the \emph{Cartan connection}, $A$, that is used to `roll' the homogeneous space without slipping or twisting around $\mathcal M$. The Cartan connection encodes the structure of the geometry on $\mathcal M$ by giving a set of rules for how $Y$ is to be rolled along $\mathcal M$. Concretely, $A$ is a $\mathfrak g$--valued 1--form with a right action on a principal $H$--bundle over $\mathcal M$, where $\mathfrak g$ is the Lie algebra of $G$. To understand how this structure encodes the geometry, it is illustrative to consider a simple example.

\begin{figure}[b]
    \begin{center}
	\includegraphics[width=0.9\textwidth]{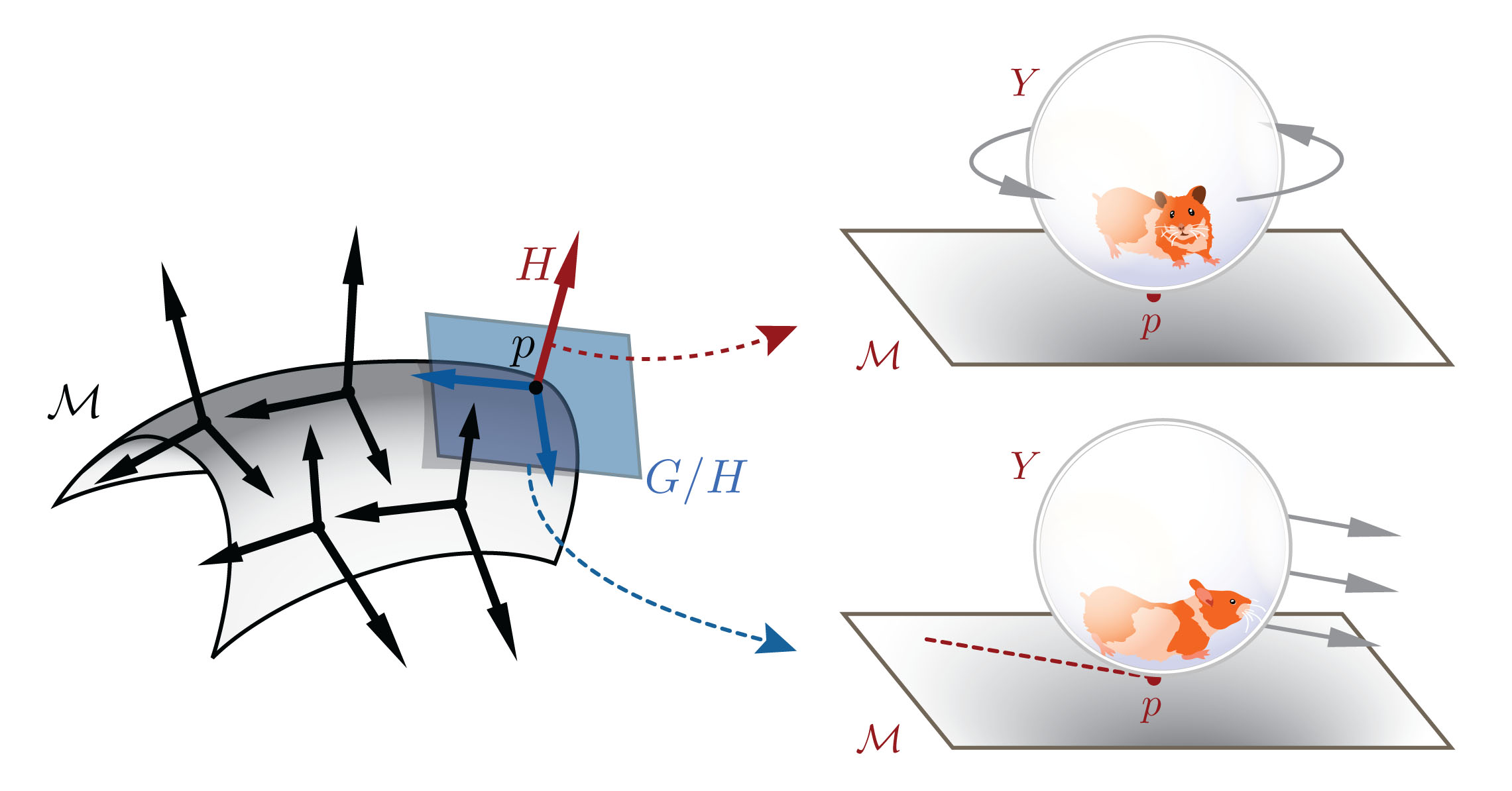}
    \caption{The hamster can move 3 different ways: one way is an $SO(2)$ rotation stabilizing $p$ (horizontal motion in the bundle). The other two ways move $p$ along independent direction in $\mathcal M$.}\label{hamsterbundle}
    \end{center}
\end{figure}

Consider the Cartan geometry formed by rolling the homogeneous space $Y = SO(3)/SO(2)$, which is a 2--sphere $\mathcal S^2$, around some closed 2d manifold. It is helpful to think, as is done in \cite{Wise:MM_paper}, of $Y$ as a hamster ball with the hamster standing above the point of contact, $p$, between the ball and the manifold as shown in figure~\ref{Hamsterball}. The information in the Cartan connection can be probed by letting the hamster move in every possible way inside the ball as all points in $\mathcal M$ are traced out. There are three different ways, shown in figure~\ref{hamsterbundle}, that the hamster can move in the ball corresponding to the three independent generators of $G=SO(3)$.
One of these is an $H=SO(2)$ rotation that stabilizes $p$. This rotation, by definition, doesn't change the point of contact between the ball and $\mathcal M$ and represents vertical motion in the bundle. The remaining two generators of $SO(3)$ represent the independent directions of $\mathcal S^2$. They are the two independent directions in which the hamster can move the point of contact between the ball and $\mathcal M$ and they correspond to horizontal motions in the bundle. The components of the Cartan connection in these directions give a natural metric on $\mathcal M$. One can think of a general Cartan geometry as being formed by rolling a generalized hamster `ball' $Y = G/H$ around an arbitrary smooth manifold $\mathcal M$. The subgroup $H$ is, then, the general group that stabilizes $G$ about the point of contact $p$ with $\mathcal M$.

In order for the connection, $A$, to correspond to a Cartan geometry, it must satisfy three requirements.\footnote{For more details on these requirements please see \cite{Sharpe:Cartan_geometry} or \cite{Wise:MM_paper}.} We will not need the details of these axioms in what follows but will describe them briefly in order to give an intuitive feel for the necessary structures of Cartan geometry. The first is essentially rolling without slipping or twisting. This says that the metric is non--degenerate so that, in particular, $Y$ and $\mathcal M$ have the same dimension. If the metric had degenerate directions, these would correspond to directions the hamster ball could slip along. The second axiom is: equivariance under $H$. This ensures that there is no preferred point in $H$. The final axiom restricts $A$ to the Maurer--Cartan form along the fibres of the bundle. 

As a last example, consider Riemannian geometry. If one identifies $G = ISO(p,q)$ and $H = SO(p,q)$, then Riemannian geometry is given by `rolling' the plane $\mathbbm R^{p,q}=ISO(p,q)/SO(p,q)$ around a manifold $\mathcal M$. Because the algebra $\mathfrak g$ decomposes as
\begin{equation}
    \mathfrak g = iso(p,q) = \mathfrak g/ \mathfrak h + \mathfrak h = \mathbbm R^{p,q} \rtimes so(p,q)
\end{equation}
the Cartan geometry is said to be \emph{reductive} and the $iso(p,q)$--valued Cartan connection, $A$, decomposes as
\begin{equation}\label{eq:red A}
    A = e + \omega,
\end{equation}
where $e$ is an $\mathbbm R^{p,q}$--valued coframe field (or metric) and $\omega$ is the usual $so(p,q)$--valued spin connection. Furthermore, the curvature of $A$
\begin{equation}
    F(A) = T(e) + R(\omega)
\end{equation}
decomposes into two pieces that depend independently on $e$ and $\omega$. The first piece, $T(e)$, depending on $e^a$, is the torsion while the second piece, $R(\omega)$, depending on $\omega^a$, is the usual Riemannian curvature. Thus, many key structures of Riemannian geometry are naturally contained in the Cartan connection $A$ and its curvature.

In what follows, we will consider two kinds of Cartan geometries. The first are the Cartan geometries modeled off the dS and AdS homogeneous spaces. These differ from those of Riemannian geometry because of the presence of a cosmological constant in the algebra of the isometry groups of these spaces. This adds an important term to the curvature, as we will see. The second kind of Cartan geometry we will consider is that modeled off the conformal sphere. These conformal geometries are importantly different from the first kind because they are \emph{non--reductive} so that the simple decomposition \eq{red A} is not possible. Fortunately, we will not need much of the structure of Cartan geometry to perform our manipulations. Nevertheless, the Cartan geometry will provide a motivation for our procedure where we reinterpret the mathematical structures of $2+1$ gravity as giving dynamic conformal geometry. The vast literature on tractor calculus suggests a potential for exploring new connections between this field of mathematics and gravity.
 
\section{$2+1$ gravity as a conformal gauge theory}

\subsection{The Chern--Simons formulation of $2+1$}\label{sec:CS}

The Einstein--Hilbert action, in $2+1$ dimensions, can be written in first order form using the Palatini action\footnote{The sign of the cosmological constant depends on our convention for the signature of the metric. We use mostly `$+$'.}
\begin{equation}\label{eq:palatini}
    S_\text{Palatini} = \int_{\mathcal M} \epsilon_{\alpha\beta\gamma} \lf( e^\alpha \wedge R^{\beta\gamma}(\omega) - \frac \Lambda 6 \, e^\alpha \wedge e^\beta \wedge e^\gamma \rt),
\end{equation}
where $e^\alpha_\mu$ is the $SO(2,1)$--invariant coframe 1--form, $R\ind{\alpha\beta}{\mu\nu}(\omega)$ is the curvature of the $SO(2,1)$--connection 1--form $\omega^{\alpha\beta}_{\mu}$, $\Lambda$ is the cosmological constant, the Newton constant, $\kappa$, is has been set to $\kappa = 2\pi$, and $\mathcal M$ is a $2+1$ dimensional smooth manifold. In our notation, Greek indices will run from $0$ to $2$ and the signature of spacetime will be $(-,+,+)$ so that the time component is $0$. Indices starting from the beginning of the alphabet represent internal indices while those starting from the middle of the alphabet will be spacetime indices and will often be suppressed in favor of differential form notation.

It was noticed in \cite{Witten:2_plus_1_cs}, that the action \eq{palatini} can be rewritten as the Chern--Simons functional of a connection 1--form, $A$, valued in $G_\Lambda=ISO(2,1)$, $SO(3,1)$, or $SO(2,2)$ depending on whether $\Lambda$ is 0, positive, or negative, respectively. This is achieved by writing the generators of $G_\Lambda$ in a basis given by
\begin{equation}
\mathbbm M_{AB} = \left(
\begin{array}{cccc}
 0 & \mp  \mathbbm J_2 & \; \mathbbm J_1 & -\ell  \; \mathbbm P_0 \\
\pm \mathbbm J_2 & 0 & \pm \mathbbm J_0 & -\ell  \; \mathbbm P_1 \\
 -\mathbbm J_1 & \mp \mathbbm J_0 & 0 & -\ell  \; \mathbbm P_2 \\
 \ell  \; \mathbbm P_0 & \ell  \; \mathbbm P_1 & \ell  \; \mathbbm P_2 & 0
\end{array}
\right) \;.
\end{equation}
where $A,B = 0\hdots 3$, $\mathbbm  P_\alpha$ is the generator of translations, $\mathbbm J_\alpha$ is the generator of $so(2,1)$ rotations\footnote{ It is related to the anti--symmetric $J_{\alpha\beta}$, by $J_\alpha = \frac 1 2 \epsilon_{\alpha\beta\gamma} J^{\alpha\beta}$.}, and $\ell$ is a constant with dimensions of length needed to make $\mathbbm M_{AB}$ dimensionless. The 4-dimensional metric that the group leaves invariant is $\eta^{AB} =\text{diag} \{-,\pm,+,+\} $, with the plus sign in the last entry in the $so(3,1)$ case, and the minus sign in the $so(2,2)$ case. In this basis, the commutation relations for the generators of $G_\Lambda$ can be compactly written, for all values of $\Lambda$, as (indices are raised with the 2+1 Minkowski metric $\eta^{\alpha\beta} = \text{diag}\{-,\pm,+\}$)
\begin{align}\label{eq:so 3 1}
    [\mathbbm J_\alpha,\mathbbm J_\beta] ~=& {\epsilon_{\alpha\beta}}^\gamma \mathbbm J_{\gamma} \,, & 
    [\mathbbm J_\alpha,\mathbbm P_\beta] ~=& {\epsilon_{\alpha\beta}}^\gamma \mathbbm P_{\gamma} \,, &
    [\mathbbm P_\alpha,\mathbbm P_\beta] ~=& \mp \frac 1 {\ell^2}~{\epsilon_{\alpha\beta}}^\gamma \mathbbm J_{\gamma} \,,
\end{align}
where ${\epsilon_{\alpha\beta}}^\gamma = \epsilon_{\alpha\beta\delta} \, \eta^{\delta\gamma}$, and the sign in the last commutator is minus for dS ($so(3,1)$, $\Lambda >0$), and plus for AdS ($so(2,2)$, $\Lambda<0$). We can relate $A$ to $e$ and $\omega$ by identifying
\begin{equation}\label{eq:A as so 3 1}
    A = e^\alpha ~ \mathbbm P_\alpha + \omega^\alpha ~ \mathbbm J_\alpha.
\end{equation}
We take $e^\alpha$ to have dimensions of length and $\omega^\alpha$ to be dimensionless so that $A$ is dimensionless. Using the commutators (\ref{eq:so 3 1}), it is straightforward to compute the curvature, $F$, of $A$
\begin{align}\label{eq:F so 3 1}
    F(A) &= dA + A\wedge A \notag\\
	 &= T^\alpha ~ \mathbbm P_\alpha + \Omega^\alpha ~ \mathbbm J_\alpha,
\end{align}
where
\begin{align}
  T^\alpha &= de^\alpha + {\epsilon^\alpha}_{\beta\gamma} \omega^{\beta} \wedge e^\gamma \\
    \Omega^\alpha &= d\omega^\alpha + \frac 1 2 {\epsilon^\alpha}_{\beta\gamma} \lf( \omega^\beta\wedge\omega^\gamma \mp \frac 1 {\ell^2}~ e^\beta\wedge e^\gamma \rt).
\end{align}
Note that the first two terms of $\Omega$ are just the usual Riemannian curvature, $R^\alpha(\omega) = d\omega^\alpha + \frac 1 2  {\epsilon^\alpha}_{\beta\gamma} \omega^\beta \wedge \omega^\gamma$ (where $R^\alpha = \frac{1}{2} {\epsilon^\alpha}_{\beta\gamma} R^{\beta\gamma}$) associated with $\omega$ while the second term is proportional to the cosmological constant.

The algebras $so(3,1)$ and $so(2,2)$ are exceptional in the sense that they admit two invariant bilinear forms: the usual Casimir defining the inner product
\begin{align}
    \mean{\mathbbm P_\alpha,\mathbbm P_\beta}_1 &= \frac {\eta_{\alpha\beta}} {\ell^2} \,, & \mean{\mathbbm J_\alpha,\mathbbm J_\beta}_1 &= \eta_{\alpha\beta}\,,
\end{align}
and an additional bilinear form defining the following inner product
\begin{equation}\label{eq:exotic trace}
    \mean{\mathbbm P_\alpha,\mathbbm J_\beta}_2 = \eta_{\alpha\beta} \,.
\end{equation}
Remarkably, the second inner product is also an invariant bilinear form for $iso(2,1)$ so it can be used for all values of $\Lambda$. That such an invariant inner product exists for $iso(2,1)$ is not guaranteed since this algebra is not semi--simple and, indeed, this convenient property does not hold in higher dimensions. If we use the trace given by the inner product \eq{exotic trace}, the Palatini action, for all values of $\Lambda$, can be written as
\begin{eqnarray}\label{eq:cs}
  S_A &=& \frac 1 2 \int_{\mathcal M} \text{Tr} \lf( A\wedge F - \frac 1 3 A\wedge A\wedge A \rt) \\
&=& \int \left( \, e_\alpha \wedge d \omega^\alpha +\frac 1 2  \epsilon_{\alpha\beta\gamma} \, e^\alpha \wedge \omega^\beta \wedge \omega^\gamma  - \frac \Lambda 6  \, \epsilon_{\alpha\beta\gamma} \, e^\alpha \wedge e^\beta \wedge e^\gamma \right)   = S_\text{Palatini} \nonumber \,,
\end{eqnarray}
by identifying
\begin{equation}
    \Lambda = \pm \frac 1 {\ell^2}.
\end{equation}
The $\pm$ distinguishes the dS $(+)$ and AdS $(-)$ cases.

That $2+1$ gravity can be cast into the form of a Chern--Simons gauge theory is well--known. That this result is not simply a trick but relies on the structures of Cartan geometry is, however, less widely recognized. The connection $A$ is indeed a Cartan connection for the Cartan geometries modeled off the homogeneous spaces $G_\Lambda/SO(2,1)$ corresponding to Minkowski space, for $\Lambda = 0$, dS space, for $\Lambda > 0$, and AdS space, for $\Lambda < 0$. These Cartan geometries are all reductive allowing for the decomposition of $A$ into the $\mathfrak g_\Lambda/so(2,1)$--invariant coframe field $e$ and the $so(2,1)$--invariant connection $\omega$. The $ \mathbbm P_a$ component of $F$, that we wrote as $T^a$, can, thus, be correctly identified as the torsion while the $ \mathbbm J_a$ component is a generalization of the usual Riemannian curvature, $R(\omega)$. The extra piece is precisely the one that leads to the cosmological constant term in the action and is present because the homogeneous spaces are non--flat. The model spaces are chosen so that they correspond to the homogeneous solutions of Einstein's equations in the presence of the appropriate cosmological constant. It is compelling that gravity results from the simplest gauge theory that one can write down in terms of a Cartan geometry locally modeled off these homogeneous spaces.

The equations of motion, as for any Chern--Simons gauge theory, enforce the vanishing of the Cartan curvature
\begin{equation}
    F = 0.
\end{equation}
The $P_a$ component of this equation is the vanishing of the torsion and is solved by requiring $\omega$ to be metric compatible. The $\mathbbm J_a$ component is the usual Einstein equation, in frame field form, for $2+1$ gravity in the presence of a cosmological constant
\begin{equation}\label{eq:EEs}
    R^\alpha -  \frac \Lambda 2 \, {\epsilon^\alpha}_{\beta\gamma} \, e^\beta \wedge e^\gamma = 0.
\end{equation}
Note that, to get the Einstein equations in metric form from \eq{EEs}, one needs to reconstruct the metric $g_{ij} = e^a_i \eta_{ab} e^b_j$ and Christoffel symbols (remembering that $\omega^{\alpha\beta} = \frac 1 2 \eps{\alpha\beta}{\gamma} \omega^\gamma$) then use the property that, in 2d, the Weyl tensor is zero. See, for instance, Carlip's book \cite{Carlip:book}.

\subsection{Hamiltonian decomposition}

For first order systems, passing to the Hamiltonian picture, following \cite{Faddeev:first_order_ham}, involves: 1) splitting the theory into space and time components and isolating all the time derivatives into a kinetic term, 2) dividing the variables into Lagrange multipliers (which enter without time derivatives in the Lagrangian) and dynamical variables, 3) for the dynamical variables, reading off the symplectic structure from the form of the action, 4) for the Lagrange multipliers, identifying the corresponding constraints and checking that they are first class using the appropriate symplectic structure, and 5) (if necessary) applying the Dirac procedure on the constraints until a first class system is obtained. To simplify this decomposition, we will assume globally hyperbolic topology $\mathcal M = \Sigma \times \mathbbm R$, where $\Sigma$ is a smooth spatial manifold. The spacetime integral then splits into an integral over the real line and an integral over $\Sigma$. The integral over $\Sigma$ is expressed in terms of the pullback, $\bar A = A|_\Sigma$, which is a 1--form on $\Sigma$, and a time component $A_0$ normal to $\Sigma$, which is a 0--form on $\Sigma$. For simplicity, we will drop the bar in what follows on $A$ for convenience and understand that it is a \emph{spatial} 1--form. From now on, wedge products will be understood to act on $\Sigma$. The details of this decomposition have been carried out in \cite{Witten:2_plus_1_cs}. The resulting canonical action corresponding to \eq{cs} is\footnote{Note that, in this notation, $\Omega^\alpha$ is the pullback of the spacetime $\Omega^\alpha$ onto $\Sigma$. Similarly for $T^\alpha$.}
\begin{equation}
    S_A = \int dt \int_\Sigma \lf( \dot e^\alpha \wedge \omega_\alpha + e^\alpha_0 \, \Omega_\alpha + \omega^\alpha_0 \, T_\alpha \rt)
\end{equation}

The symplectic structure can be read off from the first term. In components, it is given by
\begin{equation}\label{eq:3d pbs}
    \pb{ e^\alpha_i(x) }{\omega^\beta_j(y)} = \epsilon_{ij} \, \eta^{\alpha\beta} \,\delta(x,y),
\end{equation}
where small Roman indices starting from the middle of the alphabet are used for spatial indices and range from 1 to 2. The $A_0$ components do not enter with time derivatives (as is always the case in gauge theories) and so are Lagrange multipliers for the constraints
\begin{align}\label{eq:3d constraints}
    \Omega^\alpha & = 0 \,, & T^\alpha &= 0\,.
\end{align}
The fact that $A_0$ is non--dynamical is a key fact that allows us to write the dynamics of $2+1$ gravity in terms of a 2-dimensional conformal Cartan connection.

The constraints \eq{3d constraints} with the symplectic structure \eq{3d pbs}, are first class and, in fact, constitute an infinitesimal representation of $\mathfrak g_\Lambda$
\begin{equation}
\begin{array}{c}
    \{ T^\alpha (x) ,T^\beta (y) \} ~=  \, {\epsilon^{\alpha\beta}}_\gamma \, T^{\gamma}(x) \, \delta(x,y) \,,  \quad  \{ \Omega^\alpha (x) ,T^\beta (y) \} ~= {\epsilon^{\alpha\beta}}_\gamma \, \Omega^{\gamma}(x) \, \delta(x,y) \,, \vspace{12pt} \\
     \{ \Omega^\alpha (x) ,\Omega^\beta(y) \} ~= \mp\frac{1}{\ell^2} \, {\epsilon^{\alpha\beta}}_\gamma \, T^{\gamma}(x) \, \delta(x,y) \,.
\end{array}
\end{equation}
This is expected from the fact that they originate from a Chern--Simons description whose constraints always form representations of the gauge algebra (see, for instance, Carlip \cite{Carlip:2_plus_1_review}). However, because of the non--standard trace, the roles of $\Omega$ and $T$ are reversed. Indeed, as can be verified explicitly, $\Omega$ generates infinitesimal translations (even though it is the $\mathbbm  J_a$ component of $F$) on the coframe field $e^\alpha$ and the connection $\omega^\alpha$ while $T$ generates infinitesimal $SO(2,1)$ rotations (even though it is the $\mathbbm P_a$ component of $F$). Explicitly, we find
\begin{align}
 & \delta  e^\alpha =  d\rho^\alpha + {\epsilon^\alpha}_\beta\gamma \left( \omega^\beta \, \rho^\gamma + e^\beta \, \tau^\gamma \right) \,,
&  \delta \omega^\alpha = d\tau^\alpha +{\epsilon^\alpha}_\beta\gamma \left(  \omega^\beta \, \tau^\gamma \mp  \frac 1 {\ell^2}  e^\beta \rho^\gamma \right) \,, \label{ConstraintAlgebra-dS-AdS}
\end{align}
for some smearings $\rho_\alpha$ and $\tau_\alpha$, where
\begin{align}
 \delta f (x) = \int_\Sigma d^2y \pb{\rho_\beta(y)~\Omega^\beta(y) +\tau_\beta(y)~T^\beta(y)}{f(x)} \,. \label{GaugeTransf-dS-AdS}
\end{align}
We can reproduce the full transformation laws for the connection $A_\mu$ by noting that the Lagrange multipliers are arbitrary. Technically, this can be enforced by adding constraints enforcing the vanishing of their momenta. This leads to the full set of gauge transformations
\begin{equation}
\delta A_\mu = \partial_\mu u + [ A_\mu , u] \, ,
\end{equation}
where $u = \rho^\alpha ~ \mathbbm P_\alpha + \tau^\beta~ \mathbbm  J_\beta$ and the correct transformation law for $e^\alpha_0$ and $\omega^\alpha_0$ is obtained when $\rho^a = \tau^a = 0$.

%{\color{Red}
%From the Poisson brackets with the generators (\ref{ConstraintAlgebra-dS-AdS}) we
%can't deduce the transformation laws of the Lagrange multipliers $e^\alpha_0$ and $\omega^\alpha_0$, as their Poisson brackets are zero. They have to transform,
%anyway, to keep the action invariant. Their transformation rule can be
%deduced from the gauge transformation of the spacetime connection form $A_\mu$:
%
%this is, in components:
%\begin{align}
% & \delta  e^\alpha_\mu =  \partial_\mu \rho^\alpha + {\epsilon^\alpha}_\beta\gamma \left( \omega^\beta_\mu \, \rho^\gamma + e^\beta_\mu \, \tau^\gamma \right) \,,
%&  \delta \omega^\alpha_\mu = \partial_\mu \tau^\alpha +{\epsilon^\alpha}_\beta\gamma \left(  \omega^\beta_\mu \, \tau^\gamma \mp  \frac 1 {\ell^2}  e^\beta_\mu \rho^\gamma \right) \,,\label{FullConstraintAlgebra-dS-AdS}
%\end{align}
%which includes Eq.\,(\ref{ConstraintAlgebra-dS-AdS}) as the case in which $\mu \neq 0$.
%Notice that Eq.\,(\ref{GaugeTransf-dS-AdS}), in terms of the transformation parameter $u$, reads
%\begin{align}
% \delta f (x) = \int_\Sigma d^2y \pb{\text{Tr}\left(F(A) \, u\right)  }{f(x)} \,.
%\end{align}}

The translations on the internal space generated by $\Omega$ can be related to spacetime diffeomorphisms. The infinitesimal action of diffeomorphisms on the $e^\alpha$ and $\omega^\alpha$ fields is given by the action of the Lie derivative with respect to some vector field $V^\mu$
\begin{align}
& \text{\textsterling}_V e^\alpha_\mu =  V^\nu \lf( \partial_\nu e^\alpha_\mu - \partial_\mu e^\alpha_\nu \rt) + \partial_\mu (V^\nu e^\alpha_\nu) \\
&\text{\textsterling}_V \omega^\alpha_\mu =  V^\nu \lf( \partial_\nu \omega^\alpha_\mu - \partial_\mu \omega^\alpha_\nu \rt) + \partial_\mu (V^\nu \omega^\alpha_\nu),
\end{align}
were we have reinserted spacetime indices for convenience. The difference between the action of the diffeomorphisms and the action of the $\Omega$'s on the coframe field can be computed if we relate the local translation parameters $\rho^\alpha$ valued with internal indices to the spacetime vector fields $V^\mu$ through
\begin{equation}
    \rho^\alpha = V^\mu e^\alpha_\mu.
\end{equation}
The difference is
\begin{equation}
  \delta e^\alpha_\mu -  \text{\textsterling}_V e^\alpha_\mu =  V^\nu \lf( D_\mu e^\alpha_\nu - D_\nu e^\alpha_\mu \rt) - {\epsilon^\alpha}_{\beta\gamma} V^\nu \omega^\beta_\nu e^\gamma_\mu,
\end{equation}
where
\begin{equation}
    D_\mu e^\alpha_\nu = \partial_\mu e^\alpha_\nu + {\epsilon^\alpha}_{\beta\gamma} \,  \omega^\beta_\mu \, e^\gamma_\nu.
\end{equation}
Since $D_\nu e^\alpha_\mu - D_\mu e^\alpha_\nu$ is just the torsion 2--form in 3d, the first term vanishes on--shell (remember that the equations of motion are $F_{\mu\nu}=0$). The second term is a local Lorentz transformation of $e^\alpha$ generated by
\begin{equation}
    \tau^\alpha = V^\mu \omega^\alpha_\mu. \label{eq: LorentzTransfLabel}
\end{equation}
Thus, on--shell spacetime diffeomorphisms are gauge equivalent to internal translations of the coframe field. For the connection $\omega^\alpha$, a similar argument goes through. Taking the difference of the variations,
\begin{equation}
    \delta \omega^\alpha_\mu - \text{\textsterling}_V \omega^\alpha_\mu =  V^\nu \,\Omega^\alpha_{\mu\nu} - \partial_\mu (V^\nu \omega^\alpha_\nu) -
 {\epsilon^\alpha}_{\beta\gamma} V^\nu \omega^\beta_\mu \omega^\gamma_\nu   \,,
\end{equation}
we see that the first term is zero on-shell, since $ \Omega^\alpha_{\mu\nu} \approx 0 $. The second term
is a local Lorentz transformation of $\omega^\alpha_\mu$
\begin{equation}
\delta_\tau \omega^\alpha_\mu =  \partial_\mu \tau^\alpha + {\epsilon^\alpha}_{\beta\gamma}  \, \omega^\beta_\mu \, \tau^\gamma \,.
\end{equation}
with $\tau^\alpha$ given by (\ref{eq: LorentzTransfLabel}).

\subsection{ADM phase space}\label{sec:ADM phase space}

In this section, we show how it is possible to perform a gauge fixing and subsequent phase space reduction to obtain the ADM constraints in terms of the frame field formalism. This phase space reduction is naturally suggested by the form of the Chern--Simons constraints given in the previous section and the physical interpretation of the symmetries they generate. In this procedure, the spacetime symmetries of the theory play a central role. Later, in section~\ref{sec:SD phase space}, we will perform a phase space reduction similar to this one but where the conformal invariance of the theory plays a central role. It is the different interpretation of the symmetries of the theory that distinguish the ADM formalism from SD.

In the frame field formalism, the ADM phase space consists of a coframe field, $e^a_i$, and its conjugate momentum, $\omega^a_i$, which are both 1--forms on the spatial Cauchy surfaces $\Sigma$. We, thus, have to eliminate $e^0_i$ and $\omega^0_i$ from the theory. This can be accomplished by noticing that the constraint $T^a$ can be gauge fixed by the gauge fixing condition
\begin{equation}
    e^0_i = 0.
\end{equation}
Because $e^0$ is itself a canonical coordinate, this constraint is of the special form, noticed first by Dirac in \cite{Dirac:CMC_fixing}, such that the Dirac bracket for this gauge fixing reduces to the Poisson bracket on the constraint surface (this can also be seen by direct calculation). Thus, the gauge--fixed constraint $T^a = 0$ can be seen as a strong equation for $\omega^0$, the conjugate variable to $e^0$. In this gauge,
\begin{equation}
T^a = de^a - {\epsilon^a}_b \,  \omega^0 \wedge e^b   = 0
\end{equation}
is the metric compatibility condition for $\omega^0$ in terms of $e^a$. Thus, we can set $e^0=0$ everywhere and treat $\omega^0(e^a)$ as a metric compatible spin--connection for the 2d Cauchy surfaces. The remaining constraints are the Hamiltonian and the $SO(2)$ Gauss constraint,
\begin{align}
\Omega^0 &= d\omega^0 + \epsilon_{ab} \lf( \omega^a \wedge \omega^b \mp \frac 1 {\ell^2}~ e^a \wedge e^b \rt) = 0 \,, & T^0 &= \epsilon_{ab} \, \omega^a \wedge e^b = 0 \,,
\end{align}
and the 2--Diffeomorphism constraint 
\begin{align}
\Omega^a &= d\omega^a +  {\epsilon^a}_b \omega^b \wedge \omega^0 \,.
\end{align}
This explicitly reduces to the phase space and constraints of the ADM theory in terms of frame fields when $\omega^0$ is a metric compatible spin connection. The proof that these constraints are equivalent, in this form, to the usual ADM constraints in terms of metric variables can be found, for instance, in Thiemann's book \cite{Thiemann:book}.

\subsection{Conformal variables, $\Lambda \neq 0$}\label{sec:conf variables}

In this section, we will make use of the isomorphism between the conformal group and the corresponding dS or AdS group. Because our Cauchy surfaces are two dimensional, we will use the finite dimensional group analogous to the conformal group in higher dimensions. For simplicity, we will call this group the \emph{finite} conformal group, $Conf(p,q)$ (in contrast to the full conformal group in 2d which is infinite dimensional). Using this definition, we notice two key properties of the description of $2+1$ gravity given in the previous section: 1) that the pullback, $A_i$, of $A_\mu$ onto $\Sigma$ is the only dynamical variable in the theory since $A_0$ is a Lagrange multiplier, 2) there is an isomorphism between the finite conformal group in 2 dimensions with metric $\eta_{ab} = \text{diag}\{\pm,+\}$  and $G_\Lambda = SO(3,1)$ (or $G_\Lambda = SO(2,2)$ in AdS). This will allow us to rewrite $2+1$ gravity in terms of a dynamical connection for the Cartan geometry modeled off the conformal sphere $SO(3,1)/H$ (or $SO(2,2)/H$), where $H$ is the stabilizer of null rays in dS (or AdS) space. In terms of these variables, the Einstein equations reduce to constraints that generate gauge transformations of the finite conformal group. We will then perform a particular gauge fixing and subsequent phase space reduction that will lead to the constraints of SD. We won't discuss the problematic case of $\Lambda = 0$ in this section. See section~\ref{sec:zero lambda} for details.

We will now establish the isomorphism between $SO(3,1)$ (or $SO(2,2)$) and the finite conformal group in 2d, $\text{\it Conf}\,(2)$ (or $\text{\it Conf}\,(1,1)$). Pick the following basis for $\text{\it Conf}\,(2)$-$\text{\it Conf}\,(1,1)$:
\begin{equation}
    \{\mathbbm p_a, \mathbbm k_a, \mathbbm j, \mathbbm d \}
\end{equation}
and interpret $\mathbbm p_a$ as the generators of 2d translations, $\mathbbm k_a$ as the generators of special conformal transformations, $\mathbbm j$ as the generator of $SO(2)$ rotations (or $SO(1,1)$ boosts), and $\mathbbm d$ as the generator of dilatations. The commutator algebra is given by
\begin{equation}\label{eq:conf 2 algebra}
\begin{array}{c}
{[}\mathbbm j , \mathbbm d ] = 0 \;, ~~~ [ \mathbbm p_a ,  \mathbbm p_b ]Ê=  [ \mathbbm k_a ,  \mathbbm k_b ]Ê= 0\;,
\vspace{6pt} \\
Ê[\mathbbm  j , \mathbbm p_a ]Ê=  \pm {\epsilon_a}^b\; \mathbbm p_b \;, ~~~ [ \mathbbm j , \mathbbm k_a ] =  \pm {\epsilon_a}^b \; \mathbbm k_b \;,
\vspace{6pt} \\
{[}\mathbbm  d , \mathbbm p_a ]Ê=   - \mathbbm p_a  \;, ~~~ [ \mathbbm d , \mathbbm k_a ] =  \mathbbm k_a \;,
~~~ [ \mathbbm p_a ,  \mathbbm k_b ]Ê= \epsilon_{ab} \; \mathbbm j+ \eta_{ab} \; \mathbbm d \;.
\end{array}
\end{equation}
With the identifications 
\begin{align}
& \mathbbm  p_a = \frac 1 {\sqrt 2} \left( \mathbbm M_{3a} + \mathbbm M_{0a} \right) \;, 
& \mathbbm  k_b = \frac 1 {\sqrt 2} \left( \mathbbm M_{3a} - \mathbbm M_{0a} \right) \;, ~~~~
& \mathbbm d = \mathbbm M_{03} \;, 
& \mathbbm j_{ab} = \mathbbm M_{ab} \;,
\end{align}
or, in terms of the (A)dS generators,
\begin{align}\label{eq:ds gen}
&\mathbbm  p_a = \frac 1 {\sqrt 2} \left(  \ell \; \mathbbm P_a \mp {\epsilon_a}^b \; \mathbbm J_b\right) \;, 
&\mathbbm  k_a = \frac 1 {\sqrt 2} \left(  \ell \; \mathbbm P_a \pm {\epsilon_a}^b \; \mathbbm J_b \right) \;, ~~ 
&\mathbbm d = - \ell \;\mathbbm P_0 \;, ~~ 
&\mathbbm j = \pm \mathbbm J_0 \;,
\end{align}
whose inverse relations are
\begin{align}\label{eq:isomorphism}
&\mathbbm  P_a = \frac 1 {\sqrt 2 \, \ell} \left(  \mathbbm p_a + \mathbbm k_a\right) \;, 
&\mathbbm  J_a = \frac 1 {\sqrt 2} \epsilon\indices{_a^b} \left(   \mathbbm p_b- \mathbbm k_b \right) \;, ~~~~
&\mathbbm P_0 = - \frac{\mathbbm d} \ell \;, ~~ 
&\mathbbm J_0 = \pm \mathbbm j \;,
\end{align}
this algebra is isomorphic to \eq{so 3 1}. Note that with this choice of normalization, the conformal generators, $\{\mathbbm p^a, \mathbbm k^a,\mathbbm d,\mathbbm j \}$, are dimensionless. Consequently, $\ell$, and thus $\Lambda$, drops out of the algebra, although its sign is encoded in the signature of the internal metric. This is expected because the conformal algebra has no inherent scale while the dS and AdS algebras contain the corresponding dS or AdS radius. Graphically, the identification of the generators looks like
\begin{equation}
\mathbbm M_{AB} \sim \left(
\begin{array}{cccc}
 0 & \frac 1 {\sqrt 2} ( \mathbbm p_1 -  \mathbbm k_1 ) &  \frac 1 {\sqrt 2} ( \mathbbm p_2 -  \mathbbm k_2 ) &  \mathbbm d \\
\frac 1 {\sqrt 2} ( \mathbbm k_1 -  \mathbbm p_1 ) & 0 &  \mathbbm j & - \frac 1 {\sqrt 2} ( \mathbbm p_1 +  \mathbbm k_1 ) \\
 \frac 1 {\sqrt 2} ( \mathbbm k_2 -  \mathbbm p_2 )  & - \mathbbm j & 0 & - \frac 1 {\sqrt 2} ( \mathbbm p_2 +  \mathbbm k_2 ) \\
- \mathbbm d & \frac 1 {\sqrt 2} ( \mathbbm p_1 +  \mathbbm k_1 ) &  \frac 1 {\sqrt 2} ( \mathbbm p_2 +  \mathbbm k_2 ) & 0
\end{array}
\right) \;.
\end{equation}

The pullback, $A_i$, of $A_\mu$ onto $\Sigma$ can be expanded in this basis as
\begin{equation}\label{eq:A conf 2}
    A_i = E^a_i ~ \mathbbm p_a + B^a_i ~ \mathbbm k_a + \omega_i ~ \mathbbm j + \phi_i ~ \mathbbm d.
\end{equation}
We now restrict to $\Lambda>0$ since the $2+1$ splits must be performed differently in both theories. Refer to section~\ref{sec:ads cft} for details. We interpret: $E^a$ as a frame field for a conformal geometry, $B^a$ (as we will see) is its conjugate momentum, $\omega$ will become an $SO(2)$ spin connection, and $\phi$ is the Weyl vector, which will be describe shortly. Using the linear isomorphism \eq{isomorphism} and the definition \eq{A as so 3 1}, we find that
\begin{align}\label{eq:variable map}
    E^a_i &=  \frac 1 {\sqrt 2} \lf( \frac {e^a_i} \ell - {\epsilon^a}_b \, \omega^b_i  \rt) & \phi_i &= - \frac{ e^0_i } \ell \\
    B^a_i &= \frac 1 {\sqrt 2} \lf( \frac{e^a_i} \ell + {\epsilon^a}_b \, \omega^b_i \rt) & \omega_i & = \omega^0_i.\nonumber
\end{align}
The symplectic structure of the original variables can now be mapped to the new variables. The Poisson brackets \eq{3d pbs} and the isomorphism \eq{variable map} give
\begin{align}
    \pb{E^a_i(x)}{B^b_j(y)} &=   -\epsilon^{ab} \epsilon_{ij} \delta(x,y) & \pb{\phi_i(x)}{\omega_j(y)} &=  \epsilon_{ij}\delta(x,y).
\end{align}
The inverse of \eq{variable map} can easily be computed,
\begin{align}
    e^a_i &= \frac \ell {\sqrt 2} \lf(E^a_i + B^a_i \rt) & e^0_i &= - \ell \phi_i \label{eq:e and w} \\
    \omega^a_i &= \frac {\eps a b} {\sqrt 2} \lf( E^b_i - B^b_i  \rt) & \omega^0_i &= \omega_i. \nonumber
\end{align}

The curvature components can be calculated two ways: either by inserting the isomorphism \eq{isomorphism} and the relations \eq{e and w} into the original computation of $F$, \eq{F so 3 1}, or by direct computation by using the definition of $A_i$, \eq{A conf 2}, and computing $F = dA + A\wedge A$ using the algebra \eq{conf 2 algebra}. In both cases, the result is
\begin{equation}
    F = Q^a ~ \mathbbm p_a + S^a ~ \mathbbm k_a + \Omega ~ \mathbbm j + \Theta ~ \mathbbm d,
\end{equation}
where
\begin{align}
    Q^a &= \frac 1 {\sqrt 2} \lf( \frac{T^a}\ell - \eps a b \Omega^b \rt) = dE^a + E^b \wedge  \lf( {\epsilon^a}_b \, \omega + \delta^{a}_b \phi  \rt) \notag \\
    S^a &= \frac 1 {\sqrt 2} \lf( \frac{T^a}\ell + \eps a b \Omega^b \rt) = dB^a +  B^b  \wedge  \lf( {\epsilon^a}_b \, \omega - \delta^a_b \phi  \rt)  \notag \\
    \Omega &= \Omega^0 = d\omega + \epsilon_{ab} E^a \wedge B^b \notag \\
    \Theta &= - \frac {T^0} \ell = d\phi + \eta_{ab} E^a \wedge B^b. \label{eq:conf structure eqns}
\end{align}
The constraints $\Omega^\alpha = 0$ and $T^\alpha = 0$, with Lagrange multipliers $e^\alpha_0$ and $\omega^\alpha_0$, can now be rewritten in terms of the conformal variables by redefining the Lagrange multipliers
\begin{align}
    E^a_3 &= \frac 1 {\sqrt 2} \lf( \frac {e^a_3} \ell - \epsilon^{ab} \omega^b_3 \rt) & \phi_3 &= -\frac {e^0_3} \ell \\
    B^a_3 &= \frac 1 {\sqrt 2} \lf( \frac {e^a_3} \ell + \epsilon^{ab} \omega^b_3 \rt) & \omega_3 & = \omega^0_3.
\end{align}
so that the Hamiltonian reads
\begin{equation}
    H = \int d^2x \lf[ E^a_3 \,Q_a + B^a_3 \, S_a + \phi_3 \, \Theta + \omega_3 \, \Omega \rt].
\end{equation}
Since these constraints have been obtained from a linear isomorphism of a first class algebra, the new constraint algebra remains first class.

The gauge transformations generated by these constraints can be readily computed. Defining the gauge parameter
\begin{equation}
    \delta f(x)  = \int_\Sigma d^2 y \pb{ \rho_a ~ Q^a(y) + \sigma_a ~ S^a(y) + \varphi ~ \Omega(y) + \theta ~ \Theta(y)}{f(x)},
\end{equation}
we find that the gauge transformations are
\begin{align} \label{Gauge1}
    \delta E^a &= \epsilon^{ab} \lf[ - d\sigma_b + \lf( \delta^c_b \phi - \eps c b \omega \rt) \sigma_c \rt] + \theta \eps a b E^b + \varphi E^a \\ \label{Gauge2}
    \delta B^a &= \epsilon^{ab} \lf[ d\rho_b + \lf( \delta^c_b \phi + \eps c b \omega \rt) \rho_c \rt] + \theta \eps a b B^b - \varphi B^a \\ \label{Gauge3}
    \delta \omega &= d\theta - \rho_b E^b + \sigma_b B^b \\ \label{Gauge4}
    \delta \phi &= d\phi - \eps a b \lf( \rho_a E^b + \sigma_a B^b \rt).
\end{align}
The physical interpretation of these gauge transformation is useful to note. If one sets
\begin{equation}
    \rho^a = \sigma^a = \theta = 0,
\end{equation}
the gauge transformations correspond to local conformal transformations of the coframe field $E^a$, with conformal weight $+1$, and its momentum $B^a$, with conformal weight $-1$. The Weyl vector, $\phi$, transforms like a gauge field for dilations. We will describe its role in more detail shortly. Setting
\begin{equation}
    \rho^a = \sigma^a = \varphi = 0,
\end{equation}
we find that the $E^a$ fields and $B^a$ fields transform like vectors under internal rotations with the $\omega$ field transforming like a gauge field for $SO(2)$ rotations. 

We can relate a two-parameter subgroup of the gauge transformations (\ref{Gauge1}--\ref{Gauge4}) to 2 dimensional diffeomorphisms. Note that the Lie derivative of some field $ X^a_i$ along a vector field $V^i$ is given by
\begin{equation}
    \text{\textsterling}_V X^a_i = V^j \lf( \partial_i X^a_j-\partial_j X^a_i \rt) + \partial_i \lf( V^j X^a_j \rt),
\end{equation}
If we use the gauge parameters
\begin{align}
    \rho_a &= \epsilon_{ab} V^i B_i^b & \sigma_a &= -\epsilon_{ab} V^i E_i^b \\
    \varphi &= V^i \phi_i & \theta &= V^i \omega_i,
\end{align}
then the difference between an infinitesimal diffeomorphism and this gauge transformation is
\begin{align}
    \text{\textsterling}_V E^a_i - \delta E^a_i &=V^j \, Q^a_{ij} \approx 0 \,, & \text{\textsterling}_V B^a_i - \delta B^a_i &= V^j \, S^a_{ij} \approx 0 \,,\\
    \text{\textsterling}_V \phi_i - \delta \phi_i &= V^j \, \Omega_{ij} \approx 0 \,, & \text{\textsterling}_V \omega_i - \delta  \omega_i &= V^j \, \Theta_{ij} \approx 0 \,.
\end{align}

We can now start to interpret the components of $A_i$ as the structures necessary for describing a dynamic conformal geometry. As we have stated, the $E^a$ field can be interpreted as the coframe field a conformal geometry. Explicitly, we can construct the metric
\begin{equation}
    \bar g_{ij} = E^a_i \eta_{ab} E^b_j.
\end{equation}
Under dilatations parametrized by $\varphi$, $\bar g_{ab}$ transforms conformally according to
\begin{equation}
    \bar g_{ab} \to e^{2\varphi} \bar g_{ab}.
\end{equation}
Because the dilatations are a gauge symmetry of the theory, $\bar g_{ab}$ is a conformally invariant metric. Its conjugate momentum density, $\bar \pi^{ab}$ is related to the $E^a$ and $B^a$ fields by
\begin{equation}
    \bar \pi^{ij} = \frac 1 4 \lf( E^{i}_a \eta^{ab} \epsilon^{jk} \epsilon_{bc} B^c_k + (i \leftrightarrow j) \rt),
\end{equation}
where we have defined the inverse coframe field, $E^i_a$, such that
\begin{equation}
    E^i_a E^a_j = \delta^i_j.
\end{equation}
In 2 dimensions, this can be given explicitly by
\begin{equation}
    E^i_a = \frac 1 {\det E} \epsilon^{ij} \epsilon_{ab} E^b_j.
\end{equation}

We now would like to interpret the remaining components of $A$, namely $\omega$ and $\phi$. We can get a hint of their meaning by thinking of \eq{conf structure eqns} as generalized structure equations. If we identify $\omega$ with an $SO(2)$ connection, $T^a$ looks like the torsion with a modification due to the conformal part of the geometry. Indeed, $\omega$ has the right transformation properties for an $SO(2)$ connection. The extra contribution due to the Weyl vector $\phi$ can be understood by writing the vanishing of the torsion $T^a =0$ in terms of the metric $\bar g_{ab}$ and the symmetric connection
\begin{equation}
    \bar\Gamma^i_{jk} = \frac 1 2 \lf[ E^i_a \lf( \partial_j E^a_k + E^b_k \epsilon\indices{_b^a} \omega_j \rt) + (j \leftrightarrow k) \rt]
\end{equation}
The modified torsionless condition, $Q^a = 0$, is equivalent to the modified metric compatibility condition
\begin{equation}
    \bar \nabla_i \; \bar g_{jk} = \phi_i \; \bar g_{jk}.
\end{equation}
This is precisely the condition, noticed by Weyl \cite{weyl:conformal_1918,weyl:weyl_book}, required on the connection in order to preserve angles -- and not lengths -- under parallel transport. Removing part of the structure preserved under parallel transport frees up a vector field, $\phi$, in the connection, which is called the \emph{Weyl vector}.

\subsection{Shape dynamics ($\Lambda > 0$)}\label{sec:SD phase space}

In this section, we perform a phase space reduction inspired by the conformal nature of the barred metric variables that reduces to shape dynamics. The gauge fixing we will use is inspired by the one performed in \cite{Witten:conf_gravity}. It is implemented using the gauge fixing condition
\begin{equation}
    \omega = 0,
\end{equation}
which is second class with respect to $Q^a = 0$. Because $\omega$ itself is a phase space variable, the gauge fixing $\omega = 0$ is also of the special form noted in section~\ref{sec:ADM phase space}. Thus, the gauge--fixed constraint $Q^a = 0$ can be seen as a strong equation for $\phi$, the variable conjugate to $\omega$. In this gauge,
\begin{equation}\label{eq:metric compatibility}
    Q^a(\omega) =  dE^a - \delta^a_b \phi \wedge E^b = 0
\end{equation}
is the metric compatibility condition for $\phi$ in terms of $E^a$. Thus, we can set $\omega=0$ everywhere and treat $\phi(E)$ as a metric compatible spin--connection for the 2d Cauchy surfaces. The expression \eq{metric compatibility} can be easily solved of $\phi$ in 2d. The result is
\begin{equation}
    \phi_k(E) = \frac 1 {\det E} \epsilon_{ab} \epsilon^{ij} E^a_k \partial_i E^b_j.
\end{equation}
The reduced phase space consists of a coframe field $E^a$ and $B^a$. To put the remaining constraints into a more familiar form, it is instructive to re--express the phase space in terms of $E^a_i$ and its canonically conjugate variable $\mathcal A^i_a$, given in terms of $B^a_i$ as
\begin{equation}
    \mathcal A^i_a = - \epsilon^{ij} \epsilon_{ab} B^b_j.
\end{equation}
The constraint $S^a = 0$ becomes
\begin{align}
    S^a &= dB^a + \phi(E)\wedge B^a \\
        &= \epsilon^{ij} \bar\nabla_i B^a_j
\end{align}
where $\bar\nabla_i \equiv \partial_i - \phi_i(E)$ is the standard covariant derivative in 2d. Since $S^a = 0$ and $\eps a b S^b = 0$ are equivalent in 2d, we find that the above constraint reduces to $\bar\nabla_{[i} \mathcal A^{j]}_a = 0$, or, in terms of spatial indices
\begin{equation}
    E^a_i \bar\nabla_{[j} \mathcal A^{k]}_a = 0,
\end{equation}
which is the standard diffeomorphism constraint in 2d. The $\Theta = 0$ constraint, becomes
\begin{align}
    \Theta &= \eta_{ab} B^a \wedge E^b + d\phi(E) \\
           &= \eps a b \mathcal A_a^i E^b_i + d\phi(E),
\end{align}
which generates $SO(2)$ rotations of $E^a_i$ and $\mathcal A_a^i$. There is a subtle issue with the $d\phi(E)$ piece that is addressed in Appendix~(\ref{sec:dphi issue}). The last constraint is no longer the usual Hamiltonian constraint of ADM but, rather, the dilatation constraint
\begin{align}
    \Omega &= \epsilon_{ab} B^a \wedge E^b \\
	    &= E^a_i \mathcal A^i_a,
\end{align}
which acts like a conformal transformation on $E^a_i$ and $\mathcal A^i_a$ and, hence, on the barred variables. The new constraint algebra will be first class because the Dirac bracket for the gauge fixing ensures that the new constraint algebra is weakly equivalent to the old one on the gauge fixed surface. The resulting theory is precisely SD with full conformal symmetry trading in terms of $\bar g_{ij}$ and $\bar \pi^{ij}$. Remarkably, this change of variables transforms a quadratic Hamiltonian constraint into a linear dilational constraint.

%The constraint algebra now looks like \marginpar{double check these}
%\begin{equation}
%\begin{array}{c}
%\{ \Theta(x) , \Theta(y) \} = 0 \,, \qquad \{ \Theta(x) , \Omega(y) \} = 0\,, \qquad \{ S^a(x) , \Theta(y) \} =  0 \,,
%\vspace{12pt}\\
%\{ \Omega(x) , \Omega(y) \} = \delta(x,y) \, \Omega(x)  \,, \qquad  ??
%\end{array}
%\end{equation}

%{\color{Red}
%This:
%$$
%\{ S^a(x) , \Omega(y) \} =  \epsilon^{ij} \,  B^a_i \, \nabla_j \delta(x,y)  \,.
%$$
%is wrong, as I didn't consider the commutator $\{ \omega(E) , B^a \}$...
%}

\subsection{Other values of $\Lambda$}

In this section, we address complications arising for non--positive values of the cosmological constant.

\subsubsection{$\Lambda < 0$}\label{sec:ads cft}

When $\Lambda < 0$, there is a difficulty in applying the formalism presented in the previous section: the connection, $A^\alpha$, is now valued in the isometry group of AdS$^{2+1}$, $SO(2,2)$, which is isomorphic to the finite conformal group in $1+1$ dimensions, $Conf(1,1)$. Unfortunately, this is the wrong signature to represent evolving conformal geometries on $\Sigma$, which has Euclidean signature. We will now outline how this difficulty can be overcome and leave the explicit details as an exercise.

\begin{figure}[h]
\begin{center}
\includegraphics[width=0.8\textwidth]{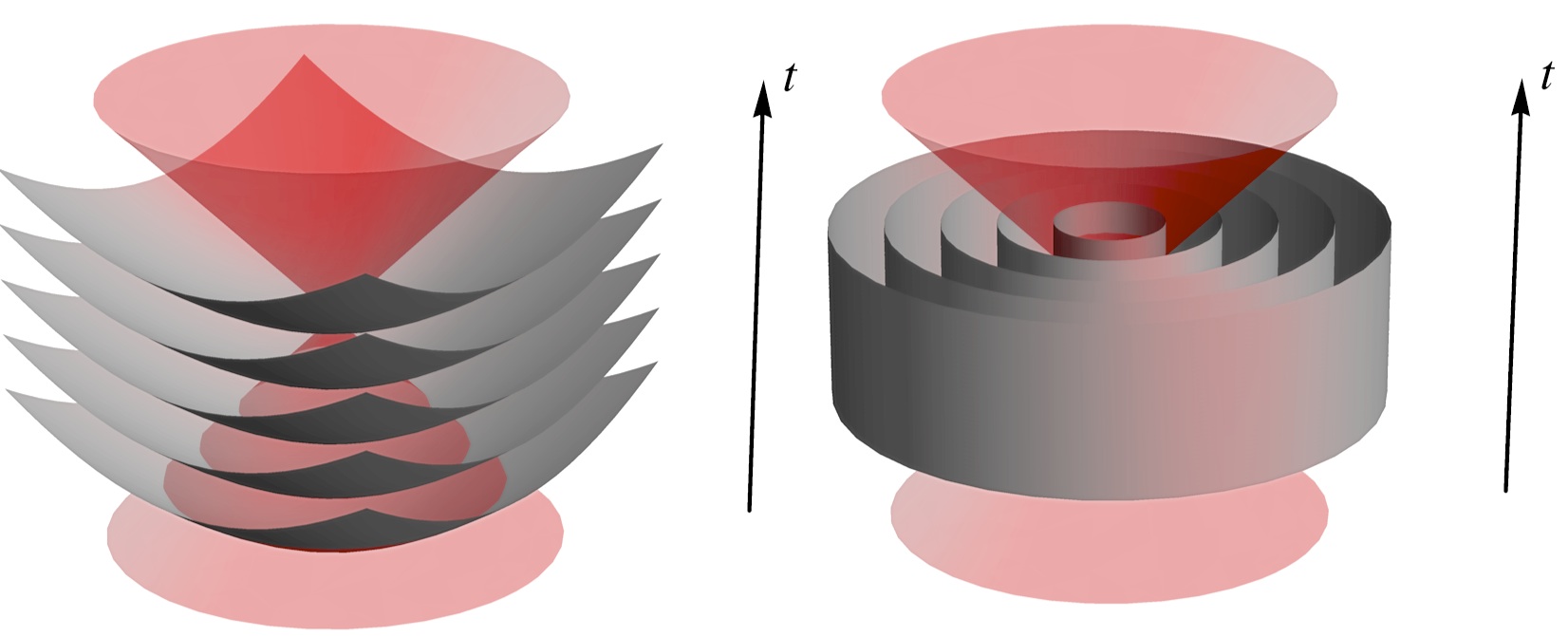}
\caption{Spacelike foliations of a portion of 2+1 dS space (left), and timelike foliation of a portion of 2+1 AdS space (right).}\label{slicings}
\end{center}
\end{figure}

The difficulty just discussed can be overcome by changing the nature of the $2+1$ decomposition. Instead of breaking up $\mathcal M$ into 2d spacelike hypersurfaces stacked in time, we can decompose $\mathcal M$ into $1+1$ dimensional timelike hypersurfaces stacked along a radial coordinate, $r$. The difference between these slicings is shown in figure~\ref{slicings}. One can use a transformation analogous to the one presented in the previous section to reinterpret the geometry of the timelike hypersurfaces as conformal $1+1$ geometries evolving radially into the bulk spacetime. This scenario is similar to what is done in the usual AdS/CFT correspondence. It is possible that the new interpretation of $A^\alpha$ as a conformal Cartan connection could be useful for studying general features of the AdS/CFT correspondence. We will not investigate this further here.

There is one additional difficulty in performing the $2+1$ split in the way described above. This difficulty arises because the timelike hypersurfaces are, in general, not Cauchy hypersurfaces. This is because the radial Hamiltonian constraint used to evolve the timelike hypersurfaces becomes hyperbolic instead of elliptic. Thus, the initial value problem is no longer well defined and one has to resort to imposing boundary conditions for the hyperbolic equations. Technically (and physically), this formalism becomes awkward. A simple way out of these difficulties is to go to the Euclidean theory. In this case, the radial Hamiltonian is elliptic and the constant--$r$ hypersurfaces are Cauchy surfaces. The connection $A^\alpha$ is then valued in $SO(3,1)$ and the only difference in the equations of the previous sections is a relative minus sign on the curvature $R(\omega)$.

\subsubsection{$\Lambda = 0$}\label{sec:zero lambda}

When $\Lambda = 0$, we can still use the non--degenerate trace given by the inner product
\begin{equation}
    \langle J_\alpha, P_\beta \rangle_2 =  \eta_{\alpha\beta}.
\end{equation}
However, the algebra $ISO(2,1)$ is no longer isometric to any conformal groups. This makes it impossible to relate the connection $A^\alpha$ to a conformal connection. Thus, the construction presented in this paper is only possible when $\Lambda \neq 0$. Strangely, the limit $\ell \to \infty$, or $\Lambda \to 0$, is still well--defined so that $\Lambda$ can be arbitrarily close to zero without creating any problems. However, at the precise value $\Lambda = 0$, the group structure changes and no interpretation in terms of conformal Cartan geometry seems possible. Fortunately, in $3+1$, $\Lambda >0$ is consistent with observations.

\section{The Shape Dynamics Hamiltonian}

The formalism presented above exactly reproduces GR in terms of conformal constraints. However, in the original formulation of SD using symmetry trading, it was convenient to single out a particular Hamiltonian as the generator of evolution. It is possible to identify this Hamiltonian in terms of our new variables. This Hamiltonian (see \cite{gryb:shape_dyn,Gomes:linking_paper} for more details) is constructed by performing the gauge fixing
\begin{equation}
    D - f(t) \sqrt g = 0,
\end{equation}
of the ADM Hamiltonian constraint, $\Omega_0$. The quantity $D = g_{ij} \pi^{ij}$ is the generator of conformal transformations. Since $f(t)$ is a spatial constant, the above gauge fixing is a way of fixing all but the global mode of the conformal factor -- or the volume. To impose this constraint, it is convenient to split the $D - f(t) \sqrt g = 0$ into the two equivalent constraints
\begin{align}\label{eq:sd gfs}
    D - \mean{D} \sqrt g &= 0 & \mean{D} - f(t) &= 0,
\end{align}
where $\mean{D} = \frac{\int D}{\int \det e} = \frac 1 V \int D$ is the \emph{mean} of $D$. The first constraint partially gauge fixes the Hamiltonian constraint by selecting the conformal factor that solves the well--known Lichnerowicz--York equation \cite{York:york_method_long}. The second constraint treats the remaining Hamiltonian as an equation for the volume, which is canonically conjugate to $\mean{D}$, the York time. This Hamiltonian, called the York Hamiltonian, gives evolution in terms of the York time.

We can use this procedure to write the SD Hamiltonian in terms of our variables. First, we have to write $D$ in terms of frame fields:
\begin{equation}
    D = \eta_{\alpha\beta} e^\alpha \wedge \omega^{\beta}.
\end{equation}
Then, the appropriate gauge fixings are given by \eq{sd gfs}. However, we need to express the frame fields in terms of the transformed variables. Using \eq{e and w}, we can express $e$ and $\omega$ in terms of $E$ and $B$. This gives, 
\begin{equation}
    D = \frac \ell 2 \lf( \det B - \det E \rt) + \ell \phi\wedge \omega,
\end{equation}
where, $\det  E = \epsilon_{ab} \epsilon^{ij}  E^a_i  E^b_j$,  $\det B = \epsilon_{ab} \epsilon^{ij}  B^a_i  B^b_j$. The constraint $D - \mean{D} \det e = 0$ should now gauge fix the Hamiltonian constraint $\Omega = d\omega + \epsilon_{ab} E^a \wedge B^b$, which becomes an equation for the variable canonically conjugate to $D- \mean{D} \det e$. This is the Lichnerowicz--York equation in terms of our new variables. Finally, the SD Hamiltonian is given by the volume (in terms of $e$)
\begin{equation}
    V = \int \det e = \frac {\ell^2} 2 \int \lf( \det E + \det B + 2 \epsilon_{ab} E^a \wedge B^b \rt)
\end{equation}
expressed as a solution to the Lichnerowicz--York equation.

Because this procedure involved solving $\Omega$ for the variable canonically conjugate to $D - \mean{D} \det e$, it is easiest in practice to first extend the phase space by introducing a parameter (called $\phi$ in \cite{gryb:shape_dyn,Gomes:linking_paper}) that parametrizes the flow of the constraint $D - \mean{D} \det e =0$. The SD Hamiltonian can then be solved in the extended theory before performing a reduction procedure. This is similar to the structure of the \emph{linking theory} used in the standard symmetry trading procedure.

It should be emphasized that this awkward procedure is not necessary using our variables. Any valid gauge fixing of the Hamiltonian constraint $\Omega$ is allowed. Since $\Omega$ simply generates conformal transformations, a condition on the conformal factor, such as
\begin{equation}
    \det B - h(t) \det E = 0
\end{equation}
would suffice.

\section{Conclusions / Discussions}

We have shown that Cartan geometry is a useful tool for understanding the connection between the spacetime picture of GR and the conformal picture of SD. Remarkably, both pictures are linked through a reinterpretation of the Cartan connection using a simple isomorphism. Originally, the equivalence between SD and GR was proved using a symmetry trading mechanism involving a phase space extension followed by a particular gauge fixing. The Cartan geometry isomorphism provides an alternative understanding of the equivalence that does not rely on the miraculous existence of the gauge fixing used in the symmetry trading procedure. Instead, there is a clear mechanism -- the isomorphism -- that adds new physical insight into the relation between SD and GR. In particular, we get as a general result that the Hamiltonian constraint, which is the generator of time translations (or timelike diffeomorphisms on--shell), can always be rewritten as the generator of conformal transformations. This is a straightforward result of the identification of $\mathbbm d = -\ell\, \mathbbm P_0$ in the isomorphism \eqref{eq:ds gen} and extends naturally to \emph{any} dimension when $\Lambda \neq 0$. We hope that by relating the isometries of dS and AdS to those of the conformal sphere, we may gain important insights into how to apply the formalism of SD to simple physical problems.

The natural extension of this work is to consider higher dimensions. We expect this extension to be complicated by the following observation: in $2+1$ dimensions, the stabilizing group, $H$, is isomorphic to the isometry group of the homogeneous space $G/H$. This results in $e$ and $\omega$ being valued in the same local algebra. In particular, the stabilizer does not appear explicitly in the Chern--Simons action. In higher dimensions, this is no longer the case and the stabilizer plays an essential role. Stelle and West \cite{Stelle_west:short,Stelle_west:long} have provided a formulation of $3+1$ gravity in terms of an $SO(4,1)$--valued connection that could be used to implement the idea presented in this paper. In this formulation, the stabilizer field appears explicitly in the action and the Palatini action is only obtained after integrating out this field. We expect this additional structure will complicate, but not ruin, the basic picture. We are currently investigating the implications of this to the $3+1$ theory.

\appendix

\section{The shape dynamics Gauss constraint} \label{sec:dphi issue}

After the phase space reduction, the $SO(2)$ Gauss constraint takes the form
\begin{equation}
     \Theta = \delta_{ab} B^a \wedge E^b + d\phi(E)
\end{equation}
instead of the obvious form
\begin{equation}
    \Theta = \delta_{ab} B^a \wedge E^b.
\end{equation}
In this section, we provide a formal argument showing that the $d\phi(E)$ piece generates a linear combination of gauge transformations on the reduced phase space. Because $\phi$ only depends upon $E$, it will only have a non--trivial action on $B$. The complication is due to the fact that the action of the eliminated constraint, $T^a = 0$, on $B$ (which was a translation on the original phase space) needs to be projected on to the reduced phase space. Because $T^a$ is the generator of a gauge transformation, the projection of its action onto a gauge fixed surface must also be a gauge transformation since no physical degrees of freedom are affected by the projection.

This can be illustrated by the following argument. Consider the Poisson bracket
\begin{equation}
    \pb{ \phi_i(x)} {B^a_j(y)} = \epsilon_{ij} \epsilon^{ab} \ddiby{ \phi_i(x) }{ E^b_j(y)}.
\end{equation}
We can compute this formally using the fact that $\phi(E)$ is defined implicitly through the solution of $T^a = 0$. Using the chain rule we find
\begin{equation}
    \ddiby{ \phi_i(x) }{ E^b_j(y)} = \int dz \lf( \ddiby{\phi_i(y)}{T^c(z)} \ddiby{T^c(z)}{E^b_j(y)} \rt)
\end{equation}
The first functional derivative can be computed by taking the inverse of $\ddiby{T}{\phi}$ which is the Poisson bracket of $T$ with $\omega$ before the phase space reduction. Thus,
\begin{equation}
    \ddiby{\phi_i(x)}{T^a(y)} = \lf[ \ddiby{T}{\phi} \rt]^{-1} = \pb{T}{\omega}^{-1}_\Gamma \equiv M_{ia}(x,y),
\end{equation}
where the subscript $\Gamma$ indicates that we have to calculate the Poisson bracket with the symplectic structure before the phase space reduction. This Poisson bracket must be invertible for us to be able to use this gauge fixing so $M_{ia}$ exists. Now, we see that
\begin{equation}
    \pb{ \phi_i(x)} {B^a_j(y)} = \int dz M_{ic}(y,z) \pb{T^c(z)}{B^a_j(y)}.
\end{equation}
This means that the flow of $B$ in the direction of $\phi(E)$ is just a linear combination of translations generated by the flow of $B$ in the direction of $T$. Thus, the $d\phi$ term can only contribute a linear combination of translations of $B$. This is really just a way of implementing the Dirac bracket for this gauge fixing.

\bibliographystyle{utphys}
\bibliography{mach}

\providecommand{\href}[2]{#2}\begingroup\raggedright\begin{thebibliography}{10}

\bibitem{Poincare:mesure_du_temps}
H.~Poincar{\'e}, ``La mesure du temps,'' {\em Revue de m{\'e}taphysique et de
  morale} {\bfseries 6} (1898) 1--13. See english translation at
  \href{http://en.wikisource.org/wiki/The_Measure_of_Time}{Wikisource}.

\bibitem{Poincare:local_time}
H.~Poincar{\'e}, ``La th{\'e}orie de Lorentz et le principe de r{\'e}action,''
  {\em Archives n{\'e}erlandaises des sciences exactes et naturelles}
  {\bfseries 5} (1900) 252--278. See
  \href{http://www.physicsinsights.org/poincare-1900.pdf}{english tranlation}.

\bibitem{Minkowski:seminal_address}
H.~Minkowski, {\em The Principle of Relativity: A Collection of Original
  Memoirs on the Special and General Theory of Relativity}, ch.~Space and Time,
  pp.~75--91.
\newblock New York: Dover, 1952.

\bibitem{barbour_el_al:physical_dof}
E.~Anderson, J.~Barbour, B.~Z. Foster, B.~Kelleher, and N.~O'Murchadha, ``{The
  physical gravitational degrees of freedom},''
  \href{http://dx.doi.org/10.1088/0264-9381/22/9/020}{{\em Class. Quant. Grav.}
  {\bfseries 22} (2005) 1795--1802},
  \href{http://arxiv.org/abs/gr-qc/0407104}{{\ttfamily arXiv:gr-qc/0407104}}.

\bibitem{gryb:shape_dyn}
H.~Gomes, S.~Gryb, and T.~Koslowski, ``{Einstein gravity as a 3D conformally
  invariant theory},''
  \href{http://dx.doi.org/10.1088/0264-9381/28/4/045005}{{\em Class. Quant.
  Grav.} {\bfseries 28} (2011) 045005},
  \href{http://arxiv.org/abs/1010.2481}{{\ttfamily arXiv:1010.2481 [gr-qc]}}.

\bibitem{Gomes:linking_paper}
H.~Gomes and T.~Koslowski, ``{The Link between General Relativity and Shape
  Dynamics},'' {\em Class.Quant.Grav.} {\bfseries 29} (2012) 075009,
  \href{http://arxiv.org/abs/1101.5974}{{\ttfamily arXiv:1101.5974 [gr-qc]}}.

\bibitem{barbour:bm_review}
J.~Barbour, ``Dynamics of pure shape, relativity and the problem of time,'' in
  {\em Decoherence and Entropy in Complex Systems}, Springer Lecture Notes in
  Physics.
\newblock 2003.
\newblock Proceedings of the Conference DICE, Piombino 2002, ed. H.-T Elze.

\bibitem{JuliansReview}
J.~Barbour, ``{Shape Dynamics. An Introduction},''
  \href{http://arxiv.org/abs/1105.0183}{{\ttfamily arXiv:1105.0183}}.

\bibitem{Sharpe:Cartan_geometry}
R.~W. Sharpe, {\em Differential Geometry: Cartan's Generalization of Klein's
  Erlangen Program}.
\newblock Springer, New York, 1997.

\bibitem{Dirac:CMC_fixing}
P.~A.~M. Dirac, ``{Fixation of coordinates in the Hamiltonian theory of
  gravitation},'' \href{http://dx.doi.org/10.1103/PhysRev.114.924}{{\em Phys.
  Rev.} {\bfseries 114} (1959) 924--930}.

\bibitem{York:cotton_tensor}
J.~J.~W. York, ``{Gravitational degrees of freedom and the initial-value
  problem},'' \href{http://dx.doi.org/10.1103/PhysRevLett.26.1656}{{\em Phys.
  Rev. Lett.} {\bfseries 26} (1971) 1656--1658}.

\bibitem{York:york_method_prl}
J.~J.~W. York, ``{Role of conformal three geometry in the dynamics of
  gravitation},'' \href{http://dx.doi.org/10.1103/PhysRevLett.28.1082}{{\em
  Phys. Rev. Lett.} {\bfseries 28} (1972) 1082--1085}.

\bibitem{Moncrief:2_plus_1_shape_space}
V.~Moncrief, ``{Reduction of the Einstein equations in (2+1)-dimensions to a
  Hamiltonian system over Teichmuller space},''
  \href{http://dx.doi.org/10.1063/1.528475}{{\em J.Math.Phys.} {\bfseries 30}
  (1989) 2907--2914}.

\bibitem{Carlip:book}
S.~Carlip, {\em Quantum gravity in 2+1 dimensions}.
\newblock Cambridge, UK: Univ. Pr., 1998.

\bibitem{York:york_method_long}
J.~J.~W. York, ``{Conformally invariant orthogonal decomposition of symmetric
  tensors on Riemannian manifolds and the initial value problem of general
  relativity},'' \href{http://dx.doi.org/10.1063/1.1666338}{{\em J. Math.
  Phys.} {\bfseries 14} (1973) 456--464}.

\bibitem{Niall_73}
N.~O'Murchadha and J.~J.~W. York, ``Existence and uniqueness of solutions of
  the Hamiltonian constraint of general relativity on compact manifolds,''
  \href{http://dx.doi.org/10.1063/1.1666225}{{\em J. Math. Phys.} {\bfseries 4}
  (1973) 1551--1557}.

\bibitem{Bodendorfer:lqg_without_Ham}
N.~Bodendorfer, A.~Stottmeister, and A.~Thurn, ``{Loop quantum gravity without
  the Hamiltonian constraint},''
  \href{http://arxiv.org/abs/1203.6525}{{\ttfamily arXiv:1203.6525 [gr-qc]}}. 5
  pages.

\bibitem{Bodendorfer:conf_coupled_sd}
N.~Bodendorfer, A.~Stottmeister, and A.~Thurn, ``{On a partially reduced phase
  space quantisation of general relativity conformally coupled to a scalar
  field},'' \href{http://arxiv.org/abs/1203.6526}{{\ttfamily arXiv:1203.6526
  [gr-qc]}}. 51 pages, 5 figures.

\bibitem{Horava:lif_point}
P.~{Ho\v rava}, ``{Quantum Gravity at a Lifshitz Point},''
  \href{http://dx.doi.org/10.1103/PhysRevD.79.084008}{{\em Phys. Rev.}
  {\bfseries D79} (2009) 084008},
  \href{http://arxiv.org/abs/0901.3775}{{\ttfamily arXiv:0901.3775 [hep-th]}}.

\bibitem{Maldacena:ads_cft}
J.~M. Maldacena, ``{The large N limit of superconformal field theories and
  supergravity},'' \href{http://dx.doi.org/10.1023/A:1026654312961}{{\em Adv.
  Theor. Math. Phys.} {\bfseries 2} (1998) 231--252},
  \href{http://arxiv.org/abs/hep-th/9711200}{{\ttfamily arXiv:hep-th/9711200}}.

\bibitem{Witten}
E.~Witten, ``{Anti-de Sitter space and holography},'' {\em Adv. Theor. Math.
  Phys.} {\bfseries 2} (1998) 253--291,
  \href{http://arxiv.org/abs/hep-th/9802150}{{\ttfamily arXiv:hep-th/9802150}}.

\bibitem{Wise:MM_paper}
D.~K. Wise, ``{MacDowell-Mansouri gravity and Cartan geometry},''
  \href{http://dx.doi.org/10.1088/0264-9381/27/15/155010}{{\em
  Class.Quant.Grav.} {\bfseries 27} (2010) 155010},
  \href{http://arxiv.org/abs/gr-qc/0611154}{{\ttfamily arXiv:gr-qc/0611154
  [gr-qc]}}.

\bibitem{Westman:Cartan_waywisers}
H.~Westman and T.~Zlosnik, ``{Gravity, Cartan geometry, and idealized
  waywisers},'' \href{http://arxiv.org/abs/1203.5709}{{\ttfamily
  arXiv:1203.5709 [gr-qc]}}.

\bibitem{Witten:2_plus_1_cs}
E.~Witten, ``{(2+1)-Dimensional Gravity as an Exactly Soluble System},''
  \href{http://dx.doi.org/10.1016/0550-3213(88)90143-5}{{\em Nucl.Phys.}
  {\bfseries B311} (1988) 46}.

\bibitem{Faddeev:first_order_ham}
L.~Faddeev and R.~Jackiw, ``{Hamiltonian Reduction of Unconstrained and
  Constrained Systems},''
  \href{http://dx.doi.org/10.1103/PhysRevLett.60.1692}{{\em Phys.Rev.Lett.}
  {\bfseries 60} (1988) 1692}.

\bibitem{Carlip:2_plus_1_review}
S.~Carlip, ``{Conformal field theory, (2+1)-dimensional gravity, and the BTZ
  black hole},'' \href{http://dx.doi.org/10.1088/0264-9381/22/12/R01}{{\em
  Class.Quant.Grav.} {\bfseries 22} (2005) R85--R124},
  \href{http://arxiv.org/abs/gr-qc/0503022}{{\ttfamily arXiv:gr-qc/0503022
  [gr-qc]}}.

\bibitem{Thiemann:book}
T.~Thiemann, {\em Modern canonical quantum general relativity}.
\newblock Cambridge Univ. Pr., 2007.
\newblock \href{http://arxiv.org/abs/gr-qc/0110034}{{\ttfamily
  arXiv:gr-qc/0110034 [gr-qc]}}.

\bibitem{weyl:conformal_1918}
H.~Weyl, ``Reine Infinitesimalgeometrie,'' {\em Math. Z.} {\bfseries 2} (1918)
  .

\bibitem{weyl:weyl_book}
H.~Weyl, {\em {Space--Time--Matter}}.
\newblock Dover Publications, fourth~ed., 1922.

\bibitem{Witten:conf_gravity}
J.~H. Horne and E.~Witten, ``{Conformal gravity in three--dimensions as a gauge
  theory},'' \href{http://dx.doi.org/10.1103/PhysRevLett.62.501}{{\em
  Phys.Rev.Lett.} {\bfseries 62} (1989) 501--504}.

\bibitem{Stelle_west:short}
K.~Stelle and P.~C. West, ``{de Sitter gauge invariance and the geometry of the
  Einstein--Cartan theory},''
  \href{http://dx.doi.org/10.1088/0305-4470/12/8/003}{{\em J.Phys.A} {\bfseries
  A12} (1979) L205--L210}.

\bibitem{Stelle_west:long}
K.~Stelle and P.~C. West, ``{Spontaneously broken de Sitter symmetry and the
  gravitational holomony group},''
  \href{http://dx.doi.org/10.1103/PhysRevD.21.1466}{{\em Phys.Rev.} {\bfseries
  D21} (1980) 1466}.

\end{thebibliography}\endgroup

\end{document}